\documentclass[12pt]{article}
\usepackage[latin9]{inputenc}
\usepackage{amsmath}
\usepackage{amssymb,color}
\usepackage{graphicx}
\usepackage{esint}
 \usepackage{cite}
\makeatletter
\usepackage{subfigure}\usepackage{amscd}
\usepackage{appendix}
\usepackage{verbatim}
\usepackage{epstopdf}

\usepackage{color}

\begin{document}
	\begin{titlepage}
		\hspace{-0.5cm}{\hbox to\hsize{\hfill{}IPMU19-0169 }}
		
		\bigskip{}
		\vspace{3\baselineskip}
		
		\begin{center}
			\textbf{\Large{Big Bounce Baryogenesis \\ }}\par
		\end{center}{\Large \par}
		
		\vspace{0.3cm}
		\begin{center}
			\textbf{
				Neil D. Barrie
			}\\
			\textbf{ }
			\par\end{center}
		
		\begin{center}
			{\it
			Kavli IPMU (WPI), UTIAS, University of Tokyo, Kashiwa, Chiba 277-8583, Japan\\
			Email: neil.barrie@ipmu.jp}\\
			\textit{\small{}}
			\par\end{center}{\small \par}
		
		\begin{center}
			\textbf{\large{}Abstract}
			\par\end{center}{\large \par}
		
		\noindent 
We explore the possibility of an Ekpyrotic contraction phase harbouring a mechanism for Baryogenesis. A Chern-Simons coupling between the fast-rolling Ekpyrotic scalar and the Standard Model Hypercharge gauge field enables the generation of a non-zero helicity during the contraction phase. The baryon number subsequently produced at the Electroweak Phase Transition is consistent with observation for a range of couplings and bounce scales. Simultaneously, the gauge field production during the contraction provides the seeds for galactic magnetic fields and sources gravitational waves, which may provide additional avenues for observational confirmation.

	\end{titlepage}
	%%%%%%%%%%%%%%%%%%%%%%%%%%%%%%%%%%%%%%%%%%%%%%%%%%%%%%%%%%%%%%%%%%%%%%%%%%%%%%%%%%%%%%%%%%%%%%%%%%%%
	\section{Introduction}

The origin of the  baryon asymmetry of the universe is one of the most important mysteries of  particle physics and cosmology.  The size of the observed baryon asymmetry is parametrized by the asymmetry parameter $\eta_{B}$ \cite{Aghanim:2018eyx},  
\begin{equation}
\eta_B = \frac{n_B}{s} \simeq 8.5\times 10^{-11} ,
\label{eta_param}
\end{equation}
where $n_B$ and $s$ are the baryon number and entropy densities of the universe, respectively.  To generate this asymmetry, in a $ \mathcal{CPT} $ conserving theory, the Sakharov conditions must be satisfied \cite{Sakharov:1967dj}. The Standard Model has all the ingredients for  producing a baryon asymmetry in the early universe but it is orders of magnitude smaller than that required to explain observations, necessitating the existence of new physics \cite{Cohen:1993nk}.

The Inflationary scenario is a well-established paradigm in standard cosmology due to its success at solving various observational problems such as the flatness, horizon, and monopole problems, as well as providing measurable predictions in the form of primordial perturbations \cite{Guth:1980zm,Linde:1981mu,Albrecht:1982wi,Mukhanov:1981xt}. Many models have been proposed and significant effort expended in the pursuit of experimental verification, but the exact mechanism for inflation is unclear \cite{Martin:2013tda}. 
%A problem which makes this more difficult is the degeneracy of the predictions of many models, and the insufficient sensitivity in current measurements of inflationary scenario observables. 
An Inflationary setting provides a unique venue in which Baryogenesis could occur, and has been an area of interest in recent years \cite{Alexander:2004us, Alexander:2011hz, Noorbala:2012fh, Barrie:2014waa, Maleknejad:2014wsa, Barrie:2015axa}. One possibility is through coupling a  pseudoscalar  inflaton, $ \phi $, to Standard Model gauge bosons through a Chern-Simons term \cite{Anber:2015yca,Jimenez:2017cdr,Domcke:2019mnd},
\begin{equation}
\frac{\phi}{4\Lambda}Y_{\mu\nu}\tilde Y^{\mu\nu}, ~~\frac{\phi}{4\Lambda}W^a_{\mu\nu}\tilde W^{a\mu\nu} ~,
\label{CS}
\end{equation}
where $ Y_{\mu\nu} $ and $ W^a_{\mu\nu} $ correspond to the Standard Model Hypercharge and Weak gauge field strength tensors, and the dual of the field strength tensor is defined as $\tilde F^{\mu\nu}= \frac{1}{2\sqrt{-g}}\epsilon^{\mu\nu\rho\sigma} F_{\rho\sigma}$.  This form  of interaction can be present in low energy effective field theories associated with a Stueckelberg field 	\cite{Stueckelberg:1900zz},  or the Green-Schwarz mechanism \cite{Green:1984sg}, with corresponding UV cut-off $ \Lambda $.  The coupling of a pseudoscalar inflaton to the Hypercharge term has been found  to generate the observed baryon asymmetry \cite{Anber:2015yca,Jimenez:2017cdr,Domcke:2019mnd}. This  mechanism provides unique connections between the cosmological background evolution and particle physics through not only Baryogenesis but  possible gravitational wave signatures through gauge field production, and the seeding of large scale magnetic fields. Thus, this mechanism provides multiple avenues for observational and experimental verification.

Despite the successes of inflationary cosmology, it suffers from many unsolved issues. These include the questions of initial conditions, fine-tuning,  the singularity problem, degeneracy of model predictions, trans-Planckian field values and violation of perturbativity \cite{Brandenberger:2016vhg}. These problems have provided motivation for considering alternative cosmologies such as Bounce Cosmologies \cite{Novello:2008ra,Lehners:2008vx,Brandenberger:2009jq,Battefeld:2014uga,Brandenberger:2016vhg,Ijjas:2018qbo}. The Bounce scenario postulates a period of space-time contraction prior to the onset of standard Big Bang cosmology, with these two epochs separated by a bounce through which the Universe transitions from a period of contraction to the usual expansion phase. One  well-studied example of a Bounce Cosmology is the Ekpyrotic universe which involves a period of ultra-slow contraction ($ \omega\gg 1 $) prior to a bounce \cite{Khoury:2001wf,Lyth:2001pf,Brandenberger:2001bs,Khoury:2001zk,Lyth:2001nv,Steinhardt:2001st,Lehners:2007ac,Buchbinder:2007ad,Buchbinder:2007tw,Cai:2012va,Ijjas:2019pyf}. A period of Ekpyrotic contraction can  be induced by a fast-rolling scalar field with a negative exponential potential, which quickly dominates the universe diluting other energy densities, including anisotropies. This model solves the flatness, horizon and monopole problems, and is capable of generating the perturbations observed in the Cosmic Microwave Background (CMB). There has been significant research and advancement in the details of this alternative cosmology model, with current observational results being in tension with the simplest models \cite{Battefeld:2014uga}. Ekpyrotic models that consist of two scalars can help to alleviate these issues, and this may  be further  resolved with improved theoretical understanding \cite{Buchbinder:2007ad,Buchbinder:2007tw}. This cosmological model provides the interesting background dynamics of a phase of cosmological contraction and has motivated investigations into applications to other open cosmological issues - the origin  of dark matter, Magnetogenesis and gravitational waves \cite{Salim:2006nw,Li:2014era,Cheung:2014nxi,Ben-Dayan:2016iks,Ito:2016fqp,Koley:2016jdw}.

%matter bounce \cite{Brandenberger:2012zb,Cai:2013kja}

In this work, we propose a Baryogenesis mechanism that takes place during an Ekpyrotic contraction phase in a Bounce Cosmology, inspired by pseudoscalar Inflation Baryogenesis scenarios \cite{Barrie:2014waa,Anber:2015yca, Barrie:2015axa,Jimenez:2017cdr,Domcke:2019mnd}. As proof of concept, a single field Ekpyrotic model will be considered, which consists of a pseudoscalar field coupled to the Standard Model Hypercharge Chern-Simons term. The fast-rolling of the pseudoscalar will lead to the generation of a net Chern-Simons number carried by the gauge fields,  providing the possibility for successful Baryogenesis. The paper will be structured as follows;  Section \ref{EKP}  describes the properties of the Ekpyrotic phase. In Section \ref{GF_evo}, the model framework and gauge field evolution will be  discussed. Section \ref{Hyp} will discuss Baryogenesis via the helicity generated in the Hypermagnetic field during the contracting phase, and other possible cosmological observables. Finally, in Section \ref{conc}, we will conclude with a discussion of the results and future directions for investigation.

\section{Ekpyrotic Contraction and Model Description}
\label{EKP}
The known issues with inflation have led to the exploration of possible alternatives to the usual inflationary paradigm, such as  string gas cosmology, bounce, and cyclic models. As with inflation, these models attempt to solve the flatness, horizon, and
monopole problems, and must be able to source the nearly scale invariant spectrum of temperature fluctuates observed in the CMB. In what follows, we will focus on a well-known type of bounce cosmology, the Ekpyrotic bounce, which solves the known cosmological problems, and can potentially resolve various issues with other bounce models, while providing the benefits over inflation of geodesic completion, and sub-Planckian field values. This type of contraction phase is a  feature of recent studies into cyclic universe models \cite{Ijjas:2019pyf}.

The period of contraction in Ekpyrotic cosmology is characterised by a large equation of state $ \omega\gg 1 $  prior to a bounce. This can be induced when the universe is dominated by a stiff form of matter such as a fast-rolling scalar field. During such a contracting phase, the stiff matter will come to dominate the total energy density of the universe,
\begin{equation}
\rho_{\textrm{Total}}=\frac{\rho_k}{a^2}+\frac{\rho_{\textrm{mat}}}{a^3}+\frac{\rho_{rad}}{a^4}+\frac{\rho_{a}}{a^6}+...+\frac{\rho_\varphi}{a^{3(1+\omega_\varphi)}}+...
\label{ener_den}
\end{equation}
where $ \rho_a $ is the energy density associated with anisotropies, and $ \rho_\varphi $ is the energy density of the fields responsible for the Ekpyrotic contraction. From Eq. (\ref{ener_den}) it is clear that in a contracting space-time background the $ \rho_\varphi $ term will quickly increase and come to dominate the energy density of the universe if $ \omega_\varphi\gg 1 $. Consequently, a sufficiently long period of $ \omega_\varphi\gg 1 $ contraction naturally leads to the suppression of any initial or generated curvature and anisotropy perturbations, while also diluting the initial radiation and matter densities. This is how the Ekpyrotic phase is able to solve the known cosmological problems, and remove  the problem of the rapid growth of initial anisotropies and any anisotropic instabilities, which can occur in other bounce scenarios.

To see how  an Ekpyrotic contracting epoch can be induced by a rolling scalar field, consider the following  relation for the equation of state parameter for a scalar $ \varphi $,
\begin{equation}
\omega=\frac{\frac{1}{2}\dot{\varphi}^2 -V(\varphi)}{\frac{1}{2}\dot{\varphi}^2 +V(\varphi)} ~.
\end{equation}
The equation of state can be $ \omega\gg 1 $ if,
\begin{equation}
 \frac{1}{2}\dot{\varphi}^2+V(\varphi) \approx 0   ~~\text{and}~~  \frac{1}{2}\dot{\varphi}^2 - V(\varphi) \gtrsim 0 ~.
 \end{equation}
 
  A simple way to achieve this, is to have the scalar $ \varphi $ fast-roll down a negative exponential potential, leading to an approximate cancellation in the denominator. That is, a scalar potential given by,
\begin{equation}
V(\varphi)\approx-V_0 e^{-\sqrt{2\epsilon}\frac{\varphi}{M_p}} ~,
\end{equation}
where the $ \epsilon $ parameter shall be referred to as the fast-roll parameter, and where $M_p=2.4\cdot 10^{18}$ GeV is the reduced Planck mass. The relation between the $ \epsilon $ and $ \omega $ is found to be,
\begin{equation}
\epsilon=\frac{3}{2}(1+\omega) ~.
\end{equation} 
 
The fast-roll parameter can be considered analogous to the inflationary slow-roll parameter where $ \epsilon_{\textrm{inf}}\ll 1 $, while instead here $ \epsilon \gg 1 $ is required with corresponding fast-roll conditions \cite{Khoury:2003rt}. Interestingly, there is a seeming duality between the Ekpyrotic and Inflationary regimes through the respective fast and slow-roll parameters, which is the motivation for the Baryogenesis mechanism considered here.

The Ekpyrotic action is of the following form,
\begin{equation}
S_\textrm{Ekp}=\int d^4 x \sqrt{-g}\left(-\frac{1}{2}M_p^2 R +\frac{1}{2}\partial_\mu \varphi\partial^\mu \varphi +V_0  e^{-\sqrt{2\epsilon}\frac{|\varphi|}{M_p}}\right)~,
\label{ekp_scalar}
\end{equation}
from which can be derived the scale factor during the Ekpyrotic Contraction, 
\begin{equation}
a= (\epsilon H_b t)^\frac{1}{\epsilon} =(\epsilon H_b  \tau)^\frac{1}{\epsilon-1}~,
\label{sca_fac} 
\end{equation}
and Hubble rate,
\begin{equation}
H= \frac{H_b}{ (\epsilon H_b|\tau|)^{\frac{\epsilon}{\epsilon-1}}} \simeq\frac{1}{\epsilon \tau}~,
\label{Hub}
\end{equation}
where we fix the bounce point to be at $ t_b=\tau_b= \frac{1}{\epsilon H_b} $ such that $ a(\tau_b)=a(t_b)=1 $, and $ t,\tau\in(-\infty,\frac{1}{\epsilon H_b} ) $ during the Ekpyrotic contraction. For large $ \epsilon $, the contraction rate is very slow such that $ t\sim \tau $. 
Through inspection of Eq. (\ref{sca_fac}) and Eq. (\ref{Hub}), it is clear that for $ \epsilon\gg 1 $ the Hubble rate can increase exponentially, while the scale factor shrinks by only an $ \mathcal{O}(1) $ factor. Thus, for $ \epsilon \sim \mathcal{O}(100) $ only a single e-fold of contraction is needed to generate 60 e-folds worth of perturbations.

In the case of $ \epsilon\gg 1 $, the equations of motion for the $ \varphi $ scalar are solved by the scaling solution \cite{Buchbinder:2007ad},
\begin{equation}
\varphi \simeq M_p \sqrt{\frac{2}{\epsilon}} \ln(-\sqrt{\epsilon V_0}\tau/M_p)~,
\label{phi_sol}
\end{equation}
and subsequently,
\begin{equation}
	\varphi^\prime\simeq\sqrt{\frac{2}{\epsilon}}\frac{M_p}{\tau}~,
	\label{scaling}
\end{equation}
which is expressed in conformal time.

Despite the advantages and simplicity of the scenario described above, the single field Ekpyrotic scenario leads to a strongly blue-tilted spectral index which is in significant tension with current CMB observations \cite{Battefeld:2014uga,Aghanim:2018eyx}.  This is one of the main issues of the original formulation of the Ekpyrotic model, that is alleviated by the introduction of an additional Ekpyrotic scalar field. If the Ekpyrotic contraction is followed by a period of kinetic dominated contraction ($ \omega=1 $) prior to the bounce, the nearly scale invariant scalar power spectrum in the CMB can be produced for $ \epsilon \sim \mathcal{O}(100) $ through the conversion of isocurvature perturbations into adiabatic perturbations by the additional scalar \cite{Ijjas:2015hcc,Levy:2015awa}. This is known as the New Ekpyrotic model, in which the background evolution is induced by two Ekpyrotic scalars and consists of a non-singular bounce sourced by a ghost condensate \cite{Buchbinder:2007ad,Lehners:2007ac,Kallosh:2007ad,Buchbinder:2007tw,Cai:2012va,Ijjas:2019pyf}. In this work we will mainly focus on the simplest single field form of the Ekpyrotic scenario, but the possibility of a two-field New Ekpyrotic scenario is parametrised through allowing an early cut-off to Chern-Simons number production prior to the bounce point, which signifies that the background evolution becomes dependent upon the second scalar.

The new Ekpyrotic scenario tends to predict relatively large non-gaussianities in the CMB, compared to inflationary models. This can provide constraints on the background evolution in combination with the scalar power spectrum. The current best constraints on the non-gaussianities, from the Planck observations \cite{Akrami:2019izv}, are,
\begin{equation}
f_{NL}^{\textrm{local}} -0.9\pm 5.1 ~,
\label{NG}
\end{equation} 
while the $ f_{NL} $ predicted by the two scalar Ekpyrotic scenario, with a period of kination prior to the bounce,  is \cite{Buchbinder:2007at,Lehners:2008my,Lehners:2010fy},
\begin{equation}
f_{NL}\propto \sqrt{\epsilon}~,
\label{NG_ekp}
\end{equation} 
where $ \epsilon \sim \mathcal{O}(100) $ can successfully produce  a nearly scale-invariant scalar power spectrum. The general form of the kinetic conversion scenario is in some tension with current observations, but can be resolved via modifications to the scalar sector. In the models presented in Ref. \cite{Fertig:2013kwa,Ijjas:2014fja}, zero non-gaussianities are generated during the Ekpyrotic contraction phase, but rather they are only produced during the conversion process prior to the bounce. This reduces the non-gaussianities to $ f_{NL}\sim \mathcal{O}(1) $ with dependence on the form of interactions between the two scalars, and efficiency of the conversion. Thus, increased  precision in measurements of the non-gaussianities alongside improvements in the theoretical understanding of the period around the bounce point and the Ekpyrotic scalar sector are necessary.

 Another characteristic of Ekpyrotic Cosmologies is that they predict a blue-tilted tensor power spectrum with a small tensor-to-scalar ratio $ r $ on CMB scales, that is below current sensitivities and difficult to measure within the near future \cite{Boyle:2003km}. The tensor perturbation spectrum is given by,
\begin{equation}
 \mathcal{P}^{v}_{T}\simeq \frac{4 k^2}{\pi^2 M_p^2}~,
 \label{Ekp_GW}
\end{equation}
where $ \epsilon \gg 1 $ has been assumed. Thus, if near future experiments such as LiteBIRD \cite{Hazumi:2019lys} are able to observe a  tensor-to-scalar ratio, significant constraints will be applied on the standard Ekpyrotic scenario. Interestingly, in the Baryogenesis scenario we consider, the fast rolling of the Ekpyrotic scalar will lead to the enhanced production of gauge fields,  via the Chern-Simons coupling, which  may lead to additional high frequency gravitational wave signatures \cite{Ben-Dayan:2016iks, Ito:2016fqp}; this will be discussed in Section \ref{Hyp}.

\section{ Gauge Field Dynamics during Contracting Phase}
\label{GF_evo}

In our model, the field $ \varphi $ will be taken to be a pseudoscalar field with the Chern-Simons coupling to the Standard Model Hypercharge field given in Eq. (\ref{CS}). The fast-rolling of $ \varphi $ induces $ \mathcal{CP} $ violating dynamics in the gauge field sector generating a non-zero Chern-Simons number density. 

 Now that $ \varphi $ is taken to be a pseudoscalar, it is necessary to reconsider the form of the potential such that it preserves $ \mathcal{P} $ transformations. One possibility is the following,
\begin{equation}
V(\varphi)= \frac{V_0}{2\cosh\left({\sqrt{2\epsilon}\frac{\varphi}{M_p}}\right)} ~,\label{cosh}
\end{equation}
which for large $ \varphi $ converges to,
\begin{equation}
V(\varphi)\approx -V_0 e^{-\sqrt{2\epsilon}\frac{|\varphi|}{M_p}}~,
\label{cosh1}
\end{equation}
 satisfying  the scaling solution provided in Eq. (\ref{phi_sol}).

In what follows, the assumption will be made that the motion of $ \varphi $ is only negligibly affected by the gauge field sector. This approximation will be justified in Section \ref{Hyp},  through the requirement of a negligible  gauge field energy density. We utilise the action in Eq. (\ref{ekp_scalar}), and the subsequent scaling solution given in Eq. (\ref{scaling}) to describe the evolution of $ \varphi $.  We require that $ \dot{\varphi}>0 $, such that the positive frequency gauge field modes are enhanced and a positive $ B $ number is generated.

We  investigate the dynamics of the pseudoscalar $ \varphi $ coupled to Chern-Simons term of the Standard Model $ U(1)_Y $ field in an Ekpyrotic contracting background.
The gauge field  Lagrangian  is given by,
\begin{equation}
\frac{1}{\sqrt{-g}}{\cal L}_g=-\frac{1}{4}g^{\mu\alpha}g^{\nu\beta}Y_{\mu\nu}Y_{\alpha\beta} -\frac{\varphi}{4\Lambda}Y_{\mu\nu}\tilde Y^{\mu\nu} ~,
\label{Lagra}
\end{equation} 
where  $Y_{\mu\nu}$  denotes the Hypercharge strength tensor, with corresponding coupling constant $ g_{1} $, and $ \Lambda $ is the UV cut-off.  The background dynamics are due to the rolling of $ \varphi $ with scale factor and Hubble rate given in Eq. (\ref{sca_fac}) and Eq. (\ref{Hub}), respectively.

 In conformal coordinates, the metric can be expressed as: $g_{\mu\nu}=a^2(\tau)\eta_{\mu\nu}$, so that the Lagrangian in Eq. (\ref{Lagra}) becomes,
\begin{equation}
{\cal L} = -\frac{1}{4}\eta^{\mu\rho}\eta^{\nu\sigma}Y_{\mu\nu}Y_{\rho\sigma} -\frac{\varphi}{8\Lambda}\epsilon^{\mu\nu\rho\sigma} Y_{\mu\nu} Y^{\rho\sigma}~.
\label{Ch3L2}
\end{equation}    

To allow analytical treatment, we will make the simplified assumption that the back-reaction on the motion of $ \varphi $ due to the production of the  gauge field $ Y_i $ is negligible.

The Lagrangian above leads to the following equation of motion for the $ Y $ gauge field,
\begin{equation}
\left(\partial_{\tau}^2-\vec \bigtriangledown^2 \right) Y^{i}+\frac{\varphi^\prime(\tau)}{\Lambda}\epsilon^{ijk}\partial_j Y_k=0~,
\label{Ch3xfieldeom}
\end{equation}
where the gauge $Y_0=\partial_i Y_i=0$ has been chosen.

The Ekpyrotic scalar motion $ \varphi^\prime $ in Eq. (\ref{Ch3xfieldeom}) is defined by the scaling solution,
\begin{equation}
\varphi^\prime =\sqrt{\frac{2}{\epsilon}}\frac{M_p}{-\tau} ~,
\end{equation}
which upon substituting into the equation of motion for the $Y_{\mu}$ gauge field gives,
\begin{equation}
\left(\partial_{\tau}^2-\vec \bigtriangledown^2 \right) Y^{ i}+\frac{2\kappa }{-\tau}\epsilon^{ijk}\partial_j Y_k=0~,
\label{Ch3xfieldeom1}
\end{equation}
where 
\begin{equation}  \kappa =\frac{M_p}{\sqrt{2\epsilon} \Lambda}~. \label{kappa}
\end{equation} 

Note, the similarity to the inflationary scenario, in which the instability parameter is defined as $ \xi=\sqrt{\frac{\epsilon_{\textrm{inf}}}{2}}\frac{ M_p}{ \Lambda} $ \cite{Anber:2015yca}.

  To  quantize this model, we promote the $ Y $ gauge boson fields to operators and assume that the boson has two possible circular polarisation states,
\begin{equation}
Y_i=\int\frac{d^3\vec k}{(2\pi)^{3/2}}\sum_{\alpha}\left[G_{\alpha}(\tau,k)\epsilon_{i\alpha}\hat a^a_{\alpha} {\rm e}^{i\vec k\cdot\vec x}+
G^{*}_{\alpha}(\tau,k)\epsilon^{*}_{i\alpha}\hat a_{\alpha}^{a\dagger}{\rm e}^{-i\vec k\cdot\vec x}
\right]~,
\label{Ch311}
\end{equation}
where $\vec \epsilon_{\pm}$ denotes the two possible helicity states of the $Y$ gauge boson ($\vec \epsilon_{+}^{*}=\vec \epsilon_{-}$) and the creation, $\hat a_{\alpha}^{\dagger}(\vec k)$, and annihilation, $\hat a_{\alpha}(\vec k)$, operators satisfy the canonical commutation relations,
\begin{equation}
\left[\hat a_{\alpha}(\vec k), \hat a_{\beta}^{\dagger}(\vec k')\right]=\delta_{\alpha\beta}\delta^3(\vec k-\vec k')~,
\end{equation}
and      
\begin{equation}
\hat a^a_{\alpha}(\vec k)\vert 0\rangle_{\tau}=0~,
\label{Ch314}
\end{equation}
where $\vert 0\rangle_{\tau}$ is an instantaneous vacuum state at time $\tau$.

The mode functions in Eq. (\ref{Ch311}) are described by the following equation, from Eq. (\ref{Ch3xfieldeom}),
\begin{equation}
G''_{\pm}+\left(k^2 \mp  \frac{2\kappa k}{-\tau}  \right)G_{\pm}=0~.
\label{Ch3modefunc}
\end{equation}

Interestingly, this wave mode function equation is equivalent to the case when $ \epsilon=3 $ and $ V(\varphi)\approx 0 $ case, which is a contracting kinetic domination epoch with $ \omega=1 $.  In this case, whether the background evolution is dictated by $ \varphi $ or not, the above wave mode equation is valid,  with $ \kappa=\frac{\varphi_b^\prime}{6 \Lambda H_b} $, where $ \varphi_b^\prime $ is the velocity of the scalar at the bounce point. If we were to consider the case of sub-dominant pseudoscalar with $ V(\varphi)=0 $, and $ \varphi'(\tau)=\varphi'_b/a(\tau)^2 $.  An upper limit on the value of $\varphi'_b$, and hence $ \kappa $, is provided by the requirement that the energy density of the pseudoscalar does not dominate the background dynamics.  Therefore, the results in Section \ref{Hyp} can be easily reinterpreted to this case.

Solving for the mode functions $ G_{\pm} $ in Eq. (\ref{Ch3modefunc}) gives,
\begin{equation}
G_{\pm}=  \frac{e^{-i k \tau}}{ \sqrt{2 k}  } e^{\pm \pi \kappa/2} U\left(\pm i \kappa, 0, 2 i k \tau\right)~,
\label{wave_mfn}
\end{equation}
where $ U $ is a Confluent Hypergeometric functions. This solution has been derived using the Wronskian normalisation and matched to $\mathcal{CP}$-invariant planewave modes at $ \tau \rightarrow -\infty $, described by,
\begin{equation}
A_{\pm}(\tau,k)=\frac{1}{\sqrt{2k}} e^{- ik\tau}~.
\label{BD}
\end{equation}

Now that we have derived the dynamics of the gauge fields during the Ekpyrotic contracting phase, it is possible to evaluate the net Chern-Simons number generated during the epoch via a Bogoliubov transformation with the adiabatic vacuum state. In the next section, we explore the production of the observed baryon asymmetry from  helical Hypermagnetic fields induced by the $ \varphi Y\tilde Y$ coupling.

	\section{Hypermagnetic Field Generation  and Baryogenesis}
	\label{Hyp}

	The possibility of Magnetogenesis and Baryogenesis having their origin in the Inflationary epoch has been studied for many years \cite{Turner:1987bw,Garretson:1992vt,Joyce:1997uy,Giovannini:1997eg,Giovannini:1999by,  Kuzmin:1985mm,	Anber:2006xt,Bamba:2006km,Martin:2007ue,Demozzi:2009fu,Durrer:2010mq,Caprini:2014mja,Adshead:2016iae,Kamada:2016eeb,Kamada:2016cnb,Caprini:2017vnn,Anber:2015yca,Jimenez:2017cdr,Domcke:2019mnd}.  For  the baryon asymmetry to be generated through the hypercharge Chern-Simons term, the helicity produced during the Ekpyrotic phase  must  survive until the Electroweak Phase Transition (EWPT). At the EWPT the large scale helical hypermagnetic fields are converted into magnetic fields and a $ B+L $ asymmetry. As the evolution of the hypermagnetic fields after reheating will be analogous to that studied in the Inflationary Baryogenesis scenario, we will follow the analysis therein \cite{Jimenez:2017cdr}.\footnote{ The weak gauge field can provide an analogous avenue for Baryogenesis, through the Chern-Simons coupling in Eq. (\ref{CS}), directly producing a non-zero $ B+L $ charge density as $ \varphi $ fast rolls. To ensure this $ n_{B+L} $ is not washed out by $ B+L $ violating sphalerons after reheating, we can introduce a single heavy right-handed neutrino that is in thermal equilibrium prior to the onset of equilibrium sphaleron processes, i.e.  $T_{\textrm{rh}}> m_{\nu_R}> T_{\textrm{sphal}}\sim 10^{12} $ GeV. The excess $ L $ of the net $ B+L $ charge is removed by the lepton violating Majorana mass term of the right-handed neutrinos, leaving a net $ B $ which is subsequently redistributed by equilibrium sphaleron processes in the known way.  Although this process is more efficient at producing the baryon asymmetry than that via the Hypercharge, the non-abelian nature of the weak gauge field will lead to additional back-reaction effects, with the linearised approximation $ g W^2\ll \partial W $  beginning to breakdown for $ \kappa\gtrsim 1 $. A more detailed analysis of this scenario is required to ensure successful Baryogenesis is possible. }

	Before proceeding to the calculating the  generated baryon asymmetry parameter and the associated magnetic fields, we find the constraint on the model parameters such that the gauge field energy density generated by the rolling of $ \varphi $ does not come to dominate the background dynamics during the Ekpyrotic phase, that is $ 3 M_p^2 H^2 \gg \rho_{CS}(\tau) $. The energy density produced by the Chern-Simons coupling at a given $ \tau $ is approximately given by,
	\begin{equation}
	\rho_{CS}^Y(\tau)=\langle 0\vert Y_{\mu\nu}\tilde Y^{\mu\nu}\vert 0\rangle\simeq  \frac{6}{\pi} C(\kappa) (\epsilon H)^4~,
	\end{equation}
	where the Hubble rate is given by Eq. (\ref{Hub}), and the integral term $ C(\kappa) $ takes the following form,
	\begin{equation}
	 C(\kappa)= \int_{0}^{2\kappa} z^2  (e^{\pi\kappa}|U(i\kappa,0,-2 i z)|^2-e^{-\pi\kappa}|U(-i\kappa,0,-2 i z)|^2) dz  ~,
	 \label{integral}
	\end{equation}
	where we will consider values of $ \kappa\geq \frac{1}{\sqrt{2\epsilon}}$, which correspond to $ \Lambda<M_p $. An approximate form of Eq. (\ref{integral}) can be determined,
	\begin{equation}
	C(\kappa)\sim 0.007 \frac{e^{2\pi\kappa}}{\kappa^4},~ \textrm{for} ~~\kappa > 1  ~,
	\end{equation}
and hence, to ensure that the dynamics of the hypermagnetic fields do not effect the background evolution induced by $ \varphi $ we require that,
	\begin{align}
	M_p&\gg   \sqrt{\frac{2C(\kappa)}{\pi}} \epsilon^2 |H_c| \nonumber \\ &
	\gg 0.07 \frac{e^{\pi\kappa}}{\kappa^2} \epsilon^2 |H_c|~,~ \textrm{for} ~~\kappa > 1 ~,
	\label{eng_con_hyp}
	\end{align}
where $ H_c $ is the Hubble rate at which gauge field production ends, for validity over the entire Ekpyrotic epoch, and we have also assumed that the number of e-folds of contraction between $ \tau_c $ and $ \tau_b $ is small. This constraint will be compared with the requirements on $ H_c,~ \epsilon $, and $ \kappa $ for successful Baryogenesis. 

We allow for the possibility that the generation of Hypermagnetic field helicity is cut-off prior to the bounce point, parametrised by $ H_c $ with $ |H_c|\leq|H_b| $.  This scenario can occur if there exists an additional Ekpyrotic scalar which begins to dominate over the original Ekpyrotic scalar, leading to a  bend in the trajectory of the scalar field space. This can then lead to a period of kinetic dominated contraction ($ \omega=1 $)  prior to the bounce, during which one of the fields $ \varphi $ quickly comes to a stop. If this is the pseudoscalar, the  time dependence of the $ \mathcal{CP} $ violating term disappears ending the gauge field production. This form of the scenario may play a role in suppressing the non-gaussianities normally produced in the New Ekpyrotic Scenario \cite{Buchbinder:2007at,Lehners:2008my,Lehners:2010fy,Fertig:2013kwa,Ijjas:2014fja}, as discussed in Section \ref{EKP}.

The Hypermagnetic helicity generated during the Ekpyrotic phase is assumed to be unchanged across the bounce point, and is matched to the end of the reheating epoch, which is taken as instantaneous and characterised by temperature $ T_{\rm rh} $. The exact dynamics of the bounce are model-dependent, and  are expected to  have only a minor effect on the generated asymmetry, as the bounce is assumed to be smooth and entropy conserving. Any such effects are parametrised in the case $  |H_c|<|H_b| $. 

We perform a Bogoliubov transformation between the wave mode functions of Eq. (\ref{wave_mfn}) and Eq. (\ref{BD}), and only consider  wave numbers that satisfy $ -k\tau <2\kappa $, the modes which contribute significantly to the asymmetry between circular polarisations. The magnetic field at the end of reheating is defined as,
	\begin{align}
	B_{\rm rh}(\tau)^2&=   \frac{1}{2\pi^2} \int_{\mu}^{2\epsilon \kappa |H_c|} k^4 (|G_{+}(\tau)|^2-|G_{-}(\tau)|^2) dk \nonumber\\ & 
	\simeq   \frac{1}{4\pi^2} (\epsilon H)^4  D(\kappa) ~,
	\label{20aB}
	\end{align}
where,
			\begin{equation}	
			D(\kappa) = \int_{0}^{2\kappa } z^3 (e^{\pi\kappa}|U(i\kappa,0,-2 i z)|^2-e^{-\pi\kappa}|U(-i\kappa,0,-2 i z)|^2) dk ~,
		\end{equation}
		which for $ \kappa>1 $ is approximately given by,
		\begin{equation}
		D(\kappa)\sim 0.015 \frac{e^{2\pi\kappa}}{\kappa^5} \sim \frac{2}{\kappa} C(\kappa)~.
		\end{equation}

Thus, the magnetic field at the end of reheating is,
	\begin{equation}	
	B_{\rm rh}(\tau_b)\simeq \frac{1}{2\pi}(\epsilon H_c)^2  \sqrt{\frac{2 C(\kappa)}{\kappa} }~, \label{B_rh}
\end{equation}
while the  correlation length of these magnetic fields can be approximated as an average of the wavelengths, according to the size of their contributions to the magnetic field energy density,
	\begin{align}
	\lambda_{\rm rh}(\tau_b)&=   2\pi \frac{\int_{\mu}^{2\epsilon \kappa |H_c|} k^3 (|G_{+}(\tau)|^2-|G_{-}(\tau)|^2) dk}{\int_{\mu}^{2\epsilon \kappa |H_c|} k^4 (|G_{+}(\tau)|^2-|G_{-}(\tau)|^2) dk} \nonumber\\ & 
	\simeq   \frac{4 \pi \kappa}{\epsilon |H_c|}   ~.
	\label{20aLB}
	\end{align}

%The sphaleron redistribution gives the following relation for the remaining baryon number, $ B=\frac{28}{79} (B-L) $ \cite{Kuzmin:1985mm}.

The detailed analysis of the evolution of the magnetic field in the thermal plasma from reheating to the EWPT is beyond the scope of this work, and as such we utilise the results from the pseudoscalar Inflationary Baryogenesis scenario \cite{Kamada:2016cnb,Jimenez:2017cdr}.
%\footnote{	Matching the inflationary and Ekpyrotic cases we can find a naive relation between the corresponding instability parameters $ \xi $ and $ \kappa $, 
%		\begin{equation*}
%		 \frac{e^{\pi\kappa/2}}{\kappa} \epsilon \sim\frac{e^{\pi \xi/2}}{\xi}
%	\end{equation*}
%		given that $ \epsilon \gg 1$ it is necessary that $ \kappa<\xi $.} 
The baryon asymmetry parameter produced at the EWPT from the magnetic field and correlation length given in Eq. (\ref{B_rh}) and (\ref{20aLB}) respectively, is,
\begin{equation}
\eta_{B}\simeq  5 \cdot 10^{-12}f(\theta_W, T_{BAU})  C(\kappa)\left(\frac{H_c}{H_b}\right)^3\left(\frac{\epsilon^2 |H_b| }{ 10^{14} ~\textrm{GeV}}\right)^{3/2}~,
\end{equation}
where $ f(\theta_W, T_{BAU}) $  parametrises the time dependence of the hypermagnetic helicity as baryon number is generated during the EWPT. There is significant uncertainty in the value  this function takes during the EWPT, with expected values lying within the range \cite{Jimenez:2017cdr},
\begin{equation} 
 5.6 \cdot 10^{-4}\lesssim f(\theta_W, T_{BAU}) \lesssim 0.32~,
 \label{frange}
\end{equation}
for the Baryogenesis temperature $ T_{BAU}\sim 135 $ GeV. Hence, the generated baryon asymmetry in our model lies in the range,
\begin{equation}
C(\kappa) \left(\frac{H_c}{H_b}\right)^3\left(\frac{\epsilon^2 |H_b| }{ 10^{17} ~\textrm{GeV}}\right)^{3/2}<\frac{\eta_{B}}{\eta_{B}^{obs}}<C(\kappa)\left(\frac{H_c}{H_b}\right)^3\left(\frac{\epsilon^2 |H_b| }{1.5 \cdot 10^{15} ~\textrm{GeV}}\right)^{3/2}~.
\label{baryo_hyp}
\end{equation}

\begin{figure}[h]
	\centering
	\includegraphics[width=0.5\textwidth]{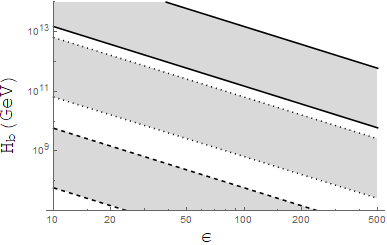}
	\caption{  Examples of the parameter regions of $ H_b $, $ \epsilon $ and $ \kappa $ that lead to successful Baryogenesis. The shaded regions subtended by the solid, dotted and dashed black lines depict the $ \kappa=1, 3,$ and 5 results respectively. The upper bounds correspond to the lower expectation of hypermagnetic field helicity to baryon number conversion, and vice versa.  }\label{constraint_hyp}
\end{figure}

The parameters required for successful Baryogenesis can be tested against the energy density constraint given in Eq. (\ref{eng_con_hyp}). First consider the constraints on  $ \epsilon^2 H_b $ from Eq. (\ref{baryo_hyp}) for the case of gauge field production until the bounce point, $ H_c=H_b $, 
\begin{equation}
\frac{1.5  \cdot 10^{15} ~\textrm{GeV}}{C(\kappa)^{2/3}} ~<\epsilon^2 |H_b| < ~\frac{ 10^{17} ~\textrm{GeV}}{C(\kappa)^{2/3}}~,
\label{baryo_hyp2}
\end{equation}
hence, considering the maximum value, Eq. (\ref{eng_con_hyp}) becomes,
	\begin{equation}
	C(\kappa)^{\frac{1}{6}}\gg  0.05 ~,
	\end{equation}
which is easily satisfied for $ \kappa>1 $. Therefore, the energy density constraint does not constrain on the parameter space when considering hypermagnetic field helicity generation up to the bounce point, and subsequent successful generation of the observed baryon asymmetry. Figure \ref{constraint_hyp}, depicts the parameter regions of successful Baryogenesis for different values of $ \kappa $. Once $ \kappa>5 $, it becomes difficult to reconcile the parameters of $ H_b $ and $ \epsilon $ with the requirements of a consistent Ekpyrotic background evolution.

\begin{figure}[h]
\begin{subfigure}
	\centering
	\includegraphics[width=0.48\textwidth]{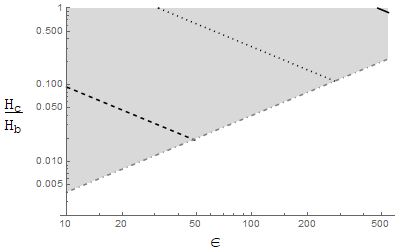}
\end{subfigure}
\begin{subfigure}
	\centering
	\includegraphics[width=0.48\textwidth]{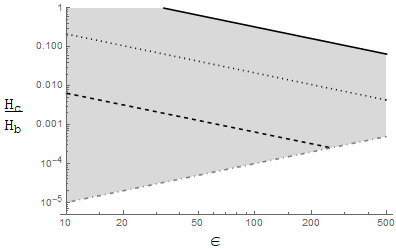}
\end{subfigure}
	\caption{   The allowed parameter space of $ H_c $, $ \epsilon $ and $ \kappa $ that leads to successful Baryogenesis and satisfies the energy density constraint, where $ T_{\rm rh} =10^{15} $ GeV and $ H_b \simeq 10^{12}$ GeV has been selected. The left-hand and right-hand plots correspond to the lower  and upper limits of Eq. (\ref{baryo_hyp}), respectively.  The dot-dashed grey line corresponds to the maximum energy density constraint given in Eq. (\ref{eng_con_hyp3}), while the upper limit on $ H_c $  is where $ H_c=H_b $. The solid, dotted and dashed black lines depict the $ \kappa=1 ,~3,$ and 5 cases respectively.  }\label{constraint}
\end{figure}

In the case of $ |H_c|<|H_b| $, there is the following constraint due to successful Baryogenesis,
\begin{equation}
\frac{1}{\epsilon C(\kappa)^{1/3}}\sqrt{\frac{1.5  \cdot 10^{15} ~\textrm{GeV}}{ |H_b|}}~<\frac{H_c}{H_b} < ~\frac{1}{\epsilon C(\kappa)^{1/3}}\sqrt{\frac{ 10^{17} ~\textrm{GeV}}{ |H_b|}}~,
\label{baryo_hyp_c}
\end{equation}
which in combination with the energy density constraint of Eq. (\ref{eng_con_hyp}) becomes the $ \kappa $-independent bound for the minimum and maximum of Eq. (\ref{frange}), respectively, 
	\begin{equation}
	\frac{H_c}{H_b}\gg \frac{\epsilon T_{\rm rh}}{2.5 \cdot 10^{20} ~\textrm{GeV}}  ~, ~~\textrm{and}~~ \frac{H_c}{H_b}\gg \frac{\epsilon T_{\rm rh}}{10^{23} ~\textrm{GeV}} ~.
	\label{eng_con_hyp2}
	\end{equation}
	
	In Figure \ref{constraint}, the  parameter space for successful Baryogenesis and consistency with Eq. (\ref{eng_con_hyp2}) is plotted, where the reheating temperature has been fixed to $ T_{\rm rh}\sim 10^{15} $ GeV. In this case, the energy density constraints used are,
		\begin{equation}
		\frac{H_c}{H_b} \gtrsim 4\cdot 10^{-4} \epsilon  ~, ~~\textrm{and}~~ \frac{H_c}{H_b}\gtrsim  10^{-6} \epsilon   ~,
		\label{eng_con_hyp3}
		\end{equation}
	which leaves significant parameter space that can produce the observed baryon asymmetry.
	
	These two scenarios generate large scale galactic magnetic fields. The evolution of the magnetic field and correlation length after the EWPT to the present day, is derived as,
	\begin{equation}
	B_p^0\simeq 2 \cdot 10^{-18} ~ \textrm{G}~ C(\kappa)^{1/3}\frac{H_c}{H_b}\left(\frac{ \epsilon^2 H_b }{  10^{13} ~\textrm{GeV}}\right)^{1/2}~,
	\end{equation}
	and
	\begin{equation}
	\lambda_p^0\simeq 6 \cdot 10^{-5}   ~\textrm{pc} ~  C(\kappa)^{1/3}\frac{H_c}{H_b}\left(\frac{\epsilon^2 H_b }{ 10^{13} ~\textrm{GeV}}\right)^{1/2}~.
	\end{equation}
	
Applying the parameters required for successful Baryogenesis given in Eq. (\ref{baryo_hyp}), the present day magnetic fields have magnitude and correlation length within the following ranges,
\begin{equation}
2.4 \cdot 10^{-17} ~ \textrm{G} <B_p^0< 2 \cdot 10^{-16} ~ \textrm{G}~,
\end{equation}
and
\begin{equation}
7 \cdot 10^{-4} ~ \textrm{pc} <\lambda_p^0< 6 \cdot 10^{-3} ~ \textrm{pc}~,
\end{equation}
with the upper and lower bound corresponding to the lower and upper limits of  Eq. (\ref{frange}), respectively. These magnetic fields can lead to interesting observational signatures due to their helical nature \cite{Tashiro:2013ita}. The magnitude of these magnetic fields are below the current upper limits, but unfortunately are too small to explain the current Blazar results. 

It should be noted that the present day magnetic fields that result from successful Baryogenesis in this Ekpyrotic mechanism are consistent with those for the Inflationary Baryogenesis scenario. Thus, observational results besides the magnetic fields and successful Baryogenesis are required to differentiate the two models.  One avenue to achieve this is through gravitational waves, through measurements of the tensor-to-scalar ratio on CMB scales, and possible higher frequency chiral gravitational wave signatures produced by the gauge field production \cite{Cook:2011hg,Ben-Dayan:2016iks}. As discussed in Section \ref{EKP}, the discovery of a  tensor-to-scalar ratio in near future CMB experiments could place significant constraints on the Ekpyrotic scenario \cite{Hazumi:2019lys}. 

The enhanced production of gauge fields is known to be able to generate unique gravitational signatures \cite{Cook:2011hg,Ben-Dayan:2016iks}, which can provide an additional avenue for observational testing of these models. To see whether this leads to observational consequences in our model, we can compare the chiral gravitational waves sourced by the Hypermagnetic field generation by the fast rolling of $ \varphi $ with those that are characteristic of Ekpyrotic Cosmologies, given in Eq. (\ref{Ekp_GW}). The gravitational waves produced by gauge field production during an Ekpyrotic contraction phase have been calculated in Ref. \cite{Ben-Dayan:2016iks}, and found to exhibit a bluer spectrum than that already predicted in Ekpyrotic Cosmology. Namely,
\begin{equation}
 \mathcal{P}^{s}_{T}\simeq 3.3 \cdot 10^{-7} \frac{e^{4\pi \kappa}}{\kappa^2 }  \frac{k^3 H_b}{M_p^4 } ~,
 \label{Ekp_GW2}
\end{equation}
for $ \epsilon\gg 1 $. Given that the two components of the gravitational wave spectrum are independent, $ \mathcal{P}^{\textrm{Total}}_{T}(k)=\mathcal{P}^{v}_{T}(k) + \mathcal{P}^{s}_{T}(k) $, it is then possible to determine when the gravitational waves sourced from the gauge field production become important. The frequency range of observational interest, for which $ \mathcal{P}^{s}_{T}(k) \ge \mathcal{P}^{v}_{T}(k) $ and successful Baryogenesis is achieved, is given by,
\begin{equation}
300 \textrm{ MHz} \left(\frac{\kappa}{5}\right)^{8/3} e^{\frac{2 \pi}{3}(5-\kappa)}>f> 31 \textrm{ kHz} \left(\frac{\epsilon}{25}\right)^{3} \left(\frac{\kappa}{5}\right)^{-2} e^{2 \pi(5-\kappa)} ~,
\label{freq_up}
\end{equation}
where the upper bound for successful Baryogenesis in Eq. ({\ref{baryo_hyp2}}) has been utilised. For the lower bound of Eq. ({\ref{baryo_hyp2}}), the frequency range of interest is bounded by,
\begin{equation}
37 \textrm{ MHz} \left(\frac{\kappa}{5}\right)^{8/3} e^{\frac{2 \pi}{3}(5-\kappa)}>f> 16 \textrm{ MHz} \left(\frac{\epsilon}{25}\right)^{3} \left(\frac{\kappa}{5}\right)^{-2} e^{2 \pi(5-\kappa)} ~,
\end{equation}
where in both scenarios the upper frequency limit is derived from the cut-off $ k\tau_b=2\kappa $, above which the effects of gauge field amplification are suppressed. For there to be significant observational consequences of the sourced gravitational waves, we require that $ \kappa $ is large,  $ \epsilon $ is suppressed, and that the  conversion of the Hypermagnetic field to baryon number at the EWPT is inefficient;  as shown in Eq. (\ref{freq_up}). In this regime, the frequencies of interest tend to be higher, and require greater sensitivities, than those probed by existing experiments. Therefore, the gauge field production can generate features in the high frequency region of the gravitational wave spectrum, which  if possible to probe in future could provide important information about the details of the Ekpyrotic mechanism. Although we have only mentioned the $ H_c=H_b $ case, similar conclusions can likely be drawn for the $ H_c<H_b $, dependent upon the details of the model.

Improved precision in the measurement of non-gaussianities and a detailed analysis of the predictions in this Ekpyrotic scenario, may play a key role in constraining the allowed parameter space for Baryogenesis. As discussed in Section \ref{EKP}, the non-gaussianities produced in Ekpyrotic Cosmology can be consistent with current observational measurements. However, in our scenario, additional non-gaussianities could be sourced by the Chern-Simons coupling between the Ekpyrotic scalar and Hypermagnetic field, possibly providing further avenues for testing the allowed parameter space. Given the dependence of this contribution on the details of the scalar sector, the period around the bounce point, and the possible utilisation of current mechanisms for reducing non-gaussianities in Ekpyrotic models, this will require a dedicated analysis that is to be completed in future work.

\section{Conclusion}
\label{conc}

 The Ekpyrotic cosmological model is able to provide solutions to the known cosmological problems as well as those associated with inflationary cosmology. We have proposed a mechanism for Baryogenesis that takes place during an Ekpyrotic contracting phase, prior to the bounce and onset of the standard radiation dominated epoch. If the evolution of the universe becomes dominated by a fast-rolling pseudoscalar with a negative exponential potential, a period of Ekpyrotic contraction can begin characterised by the equation of state $ \omega \gg 1 $.  Coupling this  pseudoscalar  to the Standard Model Hypercharge gauge group, through a  Chern-Simons term,  leads to the generation of a non-zero Chern-Simons number density prior to the  bounce point. The size of the produced Chern-Simons number density is found to explain the observed baryon asymmetry for a wide range of reasonable parameter choices. 

The large range of allowed parameters bodes well for successful Baryogenesis with a more detailed analysis required to understand the full gauge field dynamics during the contracting phase, and possible back-reaction or suppression effects. In general, the non-gaussianities generated in  Ekpyrotic models can be too large to be consistent with observation, unless a period of kination is induced by a secondary scalar field prior to the bounce. The presence of a second Ekpyrotic scalar motivated the consideration of the scenario of $ |H_c|<|H_b| $ in our analysis, in such a case the Chern-Simons number generation will end when the background trajectory becomes dominated by this secondary scalar.  This scenario gives a wide validity region in the parameter space for Baryogenesis, with  constraints derived from the requirement of sub-dominance of the gauge field energy density. 

The Hypercharge Chern-Simons coupling generates large scale magnetic fields, although the resultant magnitude and correlation length present today  is unable to explain the Blazar results and Baryogenesis simultaneously. To do this would require an extra source of suppression of the conversion of hypercharge helicity to baryon asymmetry at the EWPT. The present day magnetic field predictions in our model are consistent with those of the analogous inflationary Baryogenesis scenario, further pointing to the duality of the slow and fast-roll parameters. This makes it difficult to distinguish between the two scenarios without additional observational predictions. Some avenues for differentiation are through the tensor-to-scalar ratio, high frequency gravitational waves and non-gaussianities; with improved precision of measurements and theoretical developments being required. These scenarios warrant further investigation due to their rich phenomenology and the wide ranging implications for our understanding of the evolution of the early universe.

\section*{Acknowledgements}
This work was supported by the World Premier International Research Center Initiative (WPI), MEXT, Japan.


\begin{thebibliography}{999}
	%\cite{Aghanim:2018eyx}
	\bibitem{Aghanim:2018eyx}
	  N.~Aghanim {\it et al.} [Planck Collaboration],
	  %``Planck 2018 results. VI. Cosmological parameters,''
	  arXiv:1807.06209 [astro-ph.CO].
	  %%CITATION = ARXIV:1807.06209;%%
	  %1733 citations counted in INSPIRE as of 11 Nov 2019
	
		%\cite{Sakharov:1967dj}
		\bibitem{Sakharov:1967dj} 
		A.~D.~Sakharov,
		%``Violation of CP Invariance, c Asymmetry, and Baryon Asymmetry of the Universe,''
		Pisma Zh.\ Eksp.\ Teor.\ Fiz.\  {\bf 5}, 32 (1967).
		%%CITATION = ZFPRA,5,32;%% 
	
	%\cite{Cohen:1993nk}
	\bibitem{Cohen:1993nk}
	A.~G.~Cohen, D.~B.~Kaplan and A.~E.~Nelson,
	%``Progress in electroweak baryogenesis,''
	Ann.\ Rev.\ Nucl.\ Part.\ Sci.\  {\bf 43} (1993) 27
	doi:10.1146/annurev.ns.43.120193.000331
	[hep-ph/9302210].
	%%CITATION = doi:10.1146/annurev.ns.43.120193.000331;%%
	%855 citations counted in INSPIRE as of 03 Dec 2019

%---------Inflation------- %

	%\cite{Guth:1980zm}
	\bibitem{Guth:1980zm}
	A.~H.~Guth,
	%``The Inflationary Universe: A Possible Solution to the Horizon and Flatness Problems,''
	Phys.\ Rev.\ D {\bf 23} (1981) 347.
	%%CITATION = PHRVA,D23,347;%%
	%5204 citations counted in INSPIRE as of 11 mar 2015
	%\cite{Linde:1981mu}
	\bibitem{Linde:1981mu} 
	A.~D.~Linde,
	%``A New Inflationary Universe Scenario: A Possible Solution of the Horizon, Flatness, Homogeneity, Isotropy and Primordial Monopole Problems,''
	Phys.\ Lett.\ B {\bf 108}, 389 (1982);
	%%CITATION = PHLTA,B108,389;%%
	%2755 citations counted in INSPIRE as of 10 Jan 2014
	%\cite{Albrecht:1982wi}
	\bibitem{Albrecht:1982wi} 
	A.~Albrecht and P.~J.~Steinhardt,
	%``Cosmology for Grand Unified Theories with Radiatively Induced Symmetry Breaking,''
	Phys.\ Rev.\ Lett.\  {\bf 48}, 1220 (1982).
	%%CITATION = PRLTA,48,1220;%%
	%2535 citations counted in INSPIRE as of 10 Jan 2014 
	

	
	
	%\cite{Mukhanov:1981xt}
	\bibitem{Mukhanov:1981xt} 
	V.~F.~Mukhanov and G.~V.~Chibisov,
	%``Quantum Fluctuation and Nonsingular Universe. (In Russian),''
	JETP Lett.\  {\bf 33}, 532 (1981)
	[Pisma Zh.\ Eksp.\ Teor.\ Fiz.\  {\bf 33}, 549 (1981)].
	%%CITATION = JTPLA,33,532;%%
	%676 citations counted in INSPIRE as of 10 Jan 2014
	
%\cite{Martin:2013tda}
\bibitem{Martin:2013tda}
  J.~Martin, C.~Ringeval and V.~Vennin,
  %``Encyclopædia Inflationaris,''
  Phys.\ Dark Univ.\  {\bf 5-6} (2014) 75
  doi:10.1016/j.dark.2014.01.003
  [arXiv:1303.3787 [astro-ph.CO]].
  %%CITATION = doi:10.1016/j.dark.2014.01.003;%%
  %575 citations counted in INSPIRE as of 11 Nov 2019
	
	
%---------Chern-Simons baryogenesis models------- %


	%\cite{Alexander:2004us}
	\bibitem{Alexander:2004us} 
	S.~H.~-S.~Alexander, M.~E.~Peskin and M.~M.~Sheikh-Jabbari,
	%``Leptogenesis from gravity waves in models of inflation,''
	Phys.\ Rev.\ Lett.\  {\bf 96}, 081301 (2006)
	[hep-th/0403069];
	%%CITATION = HEP-TH/0403069;%%
	%75 citations counted in INSPIRE as of 04 Oct 2013 
	%\cite{Alexander:2011hz}
	\bibitem{Alexander:2011hz} 
	S.~Alexander, A.~Marciano and D.~Spergel,
	%``Chern-Simons Inflation and Baryogenesis,''
	JCAP {\bf 1304}, 046 (2013)
	[arXiv:1107.0318 [hep-th]]; 
	%%CITATION = ARXIV:1107.0318;%%
	%12 citations counted in INSPIRE as of 04 Oct 2013 
	%\cite{Noorbala:2012fh}
	\bibitem{Noorbala:2012fh} 
	A.~Maleknejad, M.~Noorbala and M.~M.~Sheikh-Jabbari,
	%``Inflato-Natural Leptogenesis: Leptogenesis in Chromo-Natural and Gauge Inflations,''
	arXiv:1208.2807 [hep-th];
	%%CITATION = ARXIV:1208.2807;%%
	%8 citations counted in INSPIRE as of 01 Feb 2014
	%\cite{Maleknejad:2014wsa}
	\bibitem{Maleknejad:2014wsa}
	A.~Maleknejad,
	%``Chiral Gravity Waves and Leptogenesis in Inflationary Models with non-Abelian Gauge Fields,''
	Phys.\ Rev.\ D {\bf 90} (2014) 2,  023542
	[arXiv:1401.7628 [hep-th]].
	%%CITATION = ARXIV:1401.7628;%%
	%6 citations counted in INSPIRE as of 11 mar 2015
	
		%\cite{Barrie:2014waa}
		\bibitem{Barrie:2014waa}
		N.~D.~Barrie and A.~Kobakhidze,
		%``Inflationary Baryogenesis in a Model with Gauged Baryon Number,''
		JHEP {\bf 1409} (2014) 163
		[arXiv:1401.1256 [hep-ph]].
		%%CITATION = ARXIV:1401.1256;%%
		%1 citations counted in INSPIRE as of 17 Oct 2014
		
		%\cite{Barrie:2015axa}
		\bibitem{Barrie:2015axa}
		N.~D.~Barrie and A.~Kobakhidze,
		%``Generating Luminous and Dark Matter During Inflation,''
		Mod.\ Phys.\ Lett.\ A {\bf 32} (2017) no.14,  1750087
	%	doi:10.1142/S0217732317500870
		[arXiv:1503.02366 [hep-ph]].
		%%CITATION = doi:10.1142/S0217732317500870;%%
		%4 citations counted in INSPIRE as of 19 Feb 2019
	
	
%--------Axion inflation Baryogenesis--------- %
	
%\cite{Anber:2015yca}
\bibitem{Anber:2015yca}
  M.~M.~Anber and E.~Sabancilar,
  %``Hypermagnetic Fields and Baryon Asymmetry from Pseudoscalar Inflation,''
  Phys.\ Rev.\ D {\bf 92} (2015) no.10,  101501
  doi:10.1103/PhysRevD.92.101501
  [arXiv:1507.00744 [hep-th]].
  %%CITATION = doi:10.1103/PhysRevD.92.101501;%%
  %32 citations counted in INSPIRE as of 11 Nov 2019
  
	
%\cite{Jimenez:2017cdr}
\bibitem{Jimenez:2017cdr}
  D.~Jiménez, K.~Kamada, K.~Schmitz and X.~J.~Xu,
  %``Baryon asymmetry and gravitational waves from pseudoscalar inflation,''
  JCAP {\bf 1712} (2017) 011
  doi:10.1088/1475-7516/2017/12/011
  [arXiv:1707.07943 [hep-ph]].
  %%CITATION = doi:10.1088/1475-7516/2017/12/011;%%
  %21 citations counted in INSPIRE as of 11 Nov 2019
	
	
%\cite{Domcke:2019mnd}
\bibitem{Domcke:2019mnd}
  V.~Domcke, B.~von Harling, E.~Morgante and K.~Mukaida,
  %``Baryogenesis from axion inflation,''
  JCAP {\bf 1910} (2019) no.10,  032
  doi:10.1088/1475-7516/2019/10/032
  [arXiv:1905.13318 [hep-ph]].
  %%CITATION = doi:10.1088/1475-7516/2019/10/032;%%
  %5 citations counted in INSPIRE as of 11 Nov 2019



%---------Stueckelberg------- %

\bibitem{Stueckelberg:1900zz} 
E.~C.~G.~Stueckelberg,
%``Interaction energy in electrodynamics and in the field theory of nuclear forces,''
Helv.\ Phys.\ Acta {\bf 11}, 225 (1938).
%%CITATION = HPACA,11,225;%%
%203 citations counted in INSPIRE aStueckelbergs of 17 Aug 2014

%\cite{Green:1984sg}
\bibitem{Green:1984sg}
M.~B.~Green and J.~H.~Schwarz,
%``Anomaly Cancellation in Supersymmetric D=10 Gauge Theory and Superstring Theory,''
Phys.\ Lett.\  {\bf 149B} (1984) 117.
doi:10.1016/0370-2693(84)91565-X
%%CITATION = doi:10.1016/0370-2693(84)91565-X;%%
%2733 citations counted in INSPIRE as of 03 Dec 2019



%---------Bounce review------- %
 
  %\cite{Brandenberger:2016vhg}
  \bibitem{Brandenberger:2016vhg}
    R.~Brandenberger and P.~Peter,
    %``Bouncing Cosmologies: Progress and Problems,''
    Found.\ Phys.\  {\bf 47} (2017) no.6,  797
    doi:10.1007/s10701-016-0057-0
    [arXiv:1603.05834 [hep-th]].
    %%CITATION = doi:10.1007/s10701-016-0057-0;%%
    %162 citations counted in INSPIRE as of 11 Nov 2019
    
%\cite{Novello:2008ra}
\bibitem{Novello:2008ra}
  M.~Novello and S.~E.~P.~Bergliaffa,
  %``Bouncing Cosmologies,''
  Phys.\ Rept.\  {\bf 463} (2008) 127
  doi:10.1016/j.physrep.2008.04.006
  [arXiv:0802.1634 [astro-ph]].
  %%CITATION = doi:10.1016/j.physrep.2008.04.006;%%
  %459 citations counted in INSPIRE as of 11 Nov 2019
  
%\cite{Lehners:2008vx}
\bibitem{Lehners:2008vx}
  J.~L.~Lehners,
  %``Ekpyrotic and Cyclic Cosmology,''
  Phys.\ Rept.\  {\bf 465} (2008) 223
  doi:10.1016/j.physrep.2008.06.001
  [arXiv:0806.1245 [astro-ph]].
  %%CITATION = doi:10.1016/j.physrep.2008.06.001;%%
  %214 citations counted in INSPIRE as of 11 Nov 2019
  
%\cite{Brandenberger:2009jq}
\bibitem{Brandenberger:2009jq}
  R.~H.~Brandenberger,
  %``Alternatives to the inflationary paradigm of structure formation,''
  Int.\ J.\ Mod.\ Phys.\ Conf.\ Ser.\  {\bf 01} (2011) 67
  doi:10.1142/S2010194511000109
  [arXiv:0902.4731 [hep-th]].
  %%CITATION = doi:10.1142/S2010194511000109;%%
  %103 citations counted in INSPIRE as of 12 Nov 2019
  
%\cite{Battefeld:2014uga}
\bibitem{Battefeld:2014uga}
  D.~Battefeld and P.~Peter,
  %``A Critical Review of Classical Bouncing Cosmologies,''
  Phys.\ Rept.\  {\bf 571} (2015) 1
  doi:10.1016/j.physrep.2014.12.004
  [arXiv:1406.2790 [astro-ph.CO]].
  %%CITATION = doi:10.1016/j.physrep.2014.12.004;%%
  %178 citations counted in INSPIRE as of 11 Nov 2019
 
    %\cite{Ijjas:2018qbo}
    \bibitem{Ijjas:2018qbo}
      A.~Ijjas and P.~J.~Steinhardt,
      %``Bouncing Cosmology made simple,''
      Class.\ Quant.\ Grav.\  {\bf 35} (2018) no.13,  135004
      doi:10.1088/1361-6382/aac482
      [arXiv:1803.01961 [astro-ph.CO]].
      %%CITATION = doi:10.1088/1361-6382/aac482;%%
      %15 citations counted in INSPIRE as of 11 Nov 2019
      
      
      
      
      
    
%---------Bounce and Ekp------- %

	%\cite{Khoury:2001wf}
	\bibitem{Khoury:2001wf}
	J.~Khoury, B.~A.~Ovrut, P.~J.~Steinhardt and N.~Turok,
	%``The Ekpyrotic universe: Colliding branes and the origin of the hot big bang,''
	Phys.\ Rev.\ D {\bf 64} (2001) 123522
	doi:10.1103/PhysRevD.64.123522
	[hep-th/0103239].
	%%CITATION = doi:10.1103/PhysRevD.64.123522;%%
	%1182 citations counted in INSPIRE as of 19 Jun 2019
	  
		  %\cite{Lyth:2001pf}
		  \bibitem{Lyth:2001pf}
		    D.~H.~Lyth,
		    %``The Primordial curvature perturbation in the ekpyrotic universe,''
		    Phys.\ Lett.\ B {\bf 524} (2002) 1
		    doi:10.1016/S0370-2693(01)01374-0
		    [hep-ph/0106153].
		    %%CITATION = doi:10.1016/S0370-2693(01)01374-0;%%
		    %203 citations counted in INSPIRE as of 19 Jun 2019
		    
		    
		%\cite{Khoury:2001zk}
		\bibitem{Khoury:2001zk}
		  J.~Khoury, B.~A.~Ovrut, P.~J.~Steinhardt and N.~Turok,
		  %``Density perturbations in the ekpyrotic scenario,''
		  Phys.\ Rev.\ D {\bf 66} (2002) 046005
		  doi:10.1103/PhysRevD.66.046005
		  [hep-th/0109050].
		  %%CITATION = doi:10.1103/PhysRevD.66.046005;%%
		  %286 citations counted in INSPIRE as of 19 Jun 2019
		%\cite{Lyth:2001nv}
		
	  %\cite{Brandenberger:2001bs}
	  \bibitem{Brandenberger:2001bs}
	    R.~Brandenberger and F.~Finelli,
	    %``On the spectrum of fluctuations in an effective field theory of the Ekpyrotic universe,''
	    JHEP {\bf 0111} (2001) 056
	    doi:10.1088/1126-6708/2001/11/056
	    [hep-th/0109004].
	    %%CITATION = doi:10.1088/1126-6708/2001/11/056;%%
	    %189 citations counted in INSPIRE as of 19 Jun 2019
	    
	\bibitem{Lyth:2001nv}
	  D.~H.~Lyth,
	  %``The Failure of cosmological perturbation theory in the new ekpyrotic scenario,''
	  Phys.\ Lett.\ B {\bf 526} (2002) 173
	  doi:10.1016/S0370-2693(01)01438-1
	  [hep-ph/0110007].
	  %%CITATION = doi:10.1016/S0370-2693(01)01438-1;%%
	  %103 citations counted in INSPIRE as of 19 Jun 2019
	  

	
	    
	    %\cite{Steinhardt:2001st}
	    \bibitem{Steinhardt:2001st}
	      P.~J.~Steinhardt and N.~Turok,
	      %``Cosmic evolution in a cyclic universe,''
	      Phys.\ Rev.\ D {\bf 65} (2002) 126003
	      doi:10.1103/PhysRevD.65.126003
	      [hep-th/0111098].
	      %%CITATION = doi:10.1103/PhysRevD.65.126003;%%
	      %633 citations counted in INSPIRE as of 19 Jun 2019
	      


%\cite{Lehners:2007ac}
\bibitem{Lehners:2007ac}
  J.~L.~Lehners, P.~McFadden, N.~Turok and P.~J.~Steinhardt,
  %``Generating ekpyrotic curvature perturbations before the big bang,''
  Phys.\ Rev.\ D {\bf 76} (2007) 103501
  doi:10.1103/PhysRevD.76.103501
  [hep-th/0702153 [HEP-TH]].
  %%CITATION = doi:10.1103/PhysRevD.76.103501;%%
  %181 citations counted in INSPIRE as of 11 Nov 2019


	    %\cite{Buchbinder:2007ad}
	    \bibitem{Buchbinder:2007ad}
	      E.~I.~Buchbinder, J.~Khoury and B.~A.~Ovrut,
	      %``New Ekpyrotic cosmology,''
	      Phys.\ Rev.\ D {\bf 76} (2007) 123503
	      doi:10.1103/PhysRevD.76.123503
	      [hep-th/0702154].
	      %%CITATION = doi:10.1103/PhysRevD.76.123503;%%
	      %282 citations counted in INSPIRE as of 19 Jun 2019
	      
	      %\cite{Buchbinder:2007tw}
	      \bibitem{Buchbinder:2007tw}
	        E.~I.~Buchbinder, J.~Khoury and B.~A.~Ovrut,
	        %``On the initial conditions in new ekpyrotic cosmology,''
	        JHEP {\bf 0711} (2007) 076
	        doi:10.1088/1126-6708/2007/11/076
	        [arXiv:0706.3903 [hep-th]].
	        %%CITATION = doi:10.1088/1126-6708/2007/11/076;%%
	        %81 citations counted in INSPIRE as of 11 Nov 2019
	      
	     %\cite{Cai:2012va}
	     \bibitem{Cai:2012va}
	       Y.~F.~Cai, D.~A.~Easson and R.~Brandenberger,
	       %``Towards a Nonsingular Bouncing Cosmology,''
	       JCAP {\bf 1208} (2012) 020
	       doi:10.1088/1475-7516/2012/08/020
	       [arXiv:1206.2382 [hep-th]].
	       %%CITATION = doi:10.1088/1475-7516/2012/08/020;%%
	       %234 citations counted in INSPIRE as of 11 Nov 2019
	      
	      
	      	
	      
	      %\cite{Ijjas:2019pyf}
	      \bibitem{Ijjas:2019pyf}
	      A.~Ijjas and P.~J.~Steinhardt,
	      %``A new kind of cyclic universe,''
	      arXiv:1904.08022 [gr-qc].
	      %%CITATION = ARXIV:1904.08022;%%
	      %1 citations counted in INSPIRE as of 19 Jun 2019



%-------------Usage of Ekpyrotic Phase------ %

%\cite{Salim:2006nw}
\bibitem{Salim:2006nw}
  J.~M.~Salim, N.~Souza, S.~E.~Perez Bergliaffa and T.~Prokopec,
  %``Creation of cosmological magnetic fields in a bouncing cosmology,''
  JCAP {\bf 0704} (2007) 011
  doi:10.1088/1475-7516/2007/04/011
  [astro-ph/0612281].
  %%CITATION = doi:10.1088/1475-7516/2007/04/011;%%
  %21 citations counted in INSPIRE as of 12 Nov 2019


%\cite{Li:2014era}
\bibitem{Li:2014era}
  C.~Li, R.~H.~Brandenberger and Y.~K.~E.~Cheung,
  %``Big-Bounce Genesis,''
  Phys.\ Rev.\ D {\bf 90} (2014) no.12,  123535
  doi:10.1103/PhysRevD.90.123535
  [arXiv:1403.5625 [gr-qc]].
  %%CITATION = doi:10.1103/PhysRevD.90.123535;%%
  %44 citations counted in INSPIRE as of 12 Nov 2019

%\cite{Cheung:2014nxi}
\bibitem{Cheung:2014nxi}
  Y.~K.~E.~Cheung, J.~U.~Kang and C.~Li,
  %``Dark matter in a bouncing universe,''
  JCAP {\bf 1411} (2014) 001
  doi:10.1088/1475-7516/2014/11/001
  [arXiv:1408.4387 [astro-ph.CO]].
  %%CITATION = doi:10.1088/1475-7516/2014/11/001;%%
  %16 citations counted in INSPIRE as of 12 Nov 2019
  
  %\cite{Ben-Dayan:2016iks}
  \bibitem{Ben-Dayan:2016iks}
  I.~Ben-Dayan,
  %``Gravitational Waves in Bouncing Cosmologies from Gauge Field Production,''
  JCAP {\bf 1609} (2016) 017
  doi:10.1088/1475-7516/2016/09/017
  [arXiv:1604.07899 [astro-ph.CO]].
  %%CITATION = doi:10.1088/1475-7516/2016/09/017;%%
  %20 citations counted in INSPIRE as of 20 Dec 2019
  
%\cite{Ito:2016fqp}
\bibitem{Ito:2016fqp}
  A.~Ito and J.~Soda,
  %``Primordial Gravitational Waves Induced by Magnetic Fields in an Ekpyrotic Scenario,''
  Phys.\ Lett.\ B {\bf 771} (2017) 415
  doi:10.1016/j.physletb.2017.05.017
  [arXiv:1607.07062 [hep-th]].
  %%CITATION = doi:10.1016/j.physletb.2017.05.017;%%
  %9 citations counted in INSPIRE as of 12 Nov 2019
  
  
	    %\cite{Koley:2016jdw}
	    \bibitem{Koley:2016jdw}
	      R.~Koley and S.~Samtani,
	      %``Magnetogenesis in Matter - Ekpyrotic Bouncing Cosmology,''
	      JCAP {\bf 1704} (2017) 030
	      doi:10.1088/1475-7516/2017/04/030
	      [arXiv:1612.08556 [gr-qc]].
	      %%CITATION = doi:10.1088/1475-7516/2017/04/030;%%
	      %7 citations counted in INSPIRE as of 11 Nov 2019

%-------------fast-roll------ %


%\cite{Khoury:2003rt}
\bibitem{Khoury:2003rt}
J.~Khoury, P.~J.~Steinhardt and N.~Turok,
%``Designing cyclic universe models,''
Phys.\ Rev.\ Lett.\  {\bf 92} (2004) 031302
doi:10.1103/PhysRevLett.92.031302
[hep-th/0307132].
%%CITATION = doi:10.1103/PhysRevLett.92.031302;%%
%163 citations counted in INSPIRE as of 28 Nov 2019

   

%-------------PLANCK and Ekpyrotic Phase------ %


%\cite{Ijjas:2015hcc}
\bibitem{Ijjas:2015hcc}
A.~Ijjas and P.~J.~Steinhardt,
%``Implications of Planck2015 for inflationary, ekpyrotic and anamorphic bouncing cosmologies,''
Class.\ Quant.\ Grav.\  {\bf 33} (2016) no.4,  044001
doi:10.1088/0264-9381/33/4/044001
[arXiv:1512.09010 [astro-ph.CO]].
%%CITATION = doi:10.1088/0264-9381/33/4/044001;%%
%32 citations counted in INSPIRE as of 12 Nov 2019

%-------------Scale-invariant perturbations in ekpyrotic cosmologies------ %

\bibitem{Levy:2015awa}
A.~M.~Levy, A.~Ijjas and P.~J.~Steinhardt,
%``Scale-invariant perturbations in ekpyrotic cosmologies without fine-tuning of initial conditions,''
Phys.\ Rev.\ D {\bf 92} (2015) no.6,  063524
doi:10.1103/PhysRevD.92.063524
[arXiv:1506.01011 [astro-ph.CO]].
%%CITATION = doi:10.1103/PhysRevD.92.063524;%%
%25 citations counted in INSPIRE as of 22 Nov 2019
	%-------------Ekpyrotic Ghost possible problem------ %



%\cite{Kallosh:2007ad}
\bibitem{Kallosh:2007ad}
R.~Kallosh, J.~U.~Kang, A.~D.~Linde and V.~Mukhanov,
%``The New ekpyrotic ghost,''
JCAP {\bf 0804} (2008) 018
doi:10.1088/1475-7516/2008/04/018
[arXiv:0712.2040 [hep-th]].
%%CITATION = doi:10.1088/1475-7516/2008/04/018;%%
%54 citations counted in INSPIRE as of 22 Nov 2019


  
  


%-------------Ekpyrotic Non-Gaussianity------ %


\bibitem{Akrami:2019izv}
Y.~Akrami {\it et al.} [Planck Collaboration],
%``Planck 2018 results. IX. Constraints on primordial non-Gaussianity,''
arXiv:1905.05697 [astro-ph.CO].
%%CITATION = ARXIV:1905.05697;%%
%71 citations counted in INSPIRE as of 07 Jan 2020


	      %\cite{Buchbinder:2007at}
	      \bibitem{Buchbinder:2007at}
	        E.~I.~Buchbinder, J.~Khoury and B.~A.~Ovrut,
	        %``Non-Gaussianities in new ekpyrotic cosmology,''
	        Phys.\ Rev.\ Lett.\  {\bf 100} (2008) 171302
	        doi:10.1103/PhysRevLett.100.171302
	        [arXiv:0710.5172 [hep-th]].
	        %%CITATION = doi:10.1103/PhysRevLett.100.171302;%%
	        %141 citations counted in INSPIRE as of 19 Jun 2019

%\cite{Lehners:2008my}
\bibitem{Lehners:2008my}
J.~L.~Lehners and P.~J.~Steinhardt,
%``Intuitive understanding of non-gaussianity in ekpyrotic and cyclic models,''
Phys.\ Rev.\ D {\bf 78} (2008) 023506
Erratum: [Phys.\ Rev.\ D {\bf 79} (2009) 129902]
doi:10.1103/PhysRevD.78.023506, 10.1103/PhysRevD.79.129902
[arXiv:0804.1293 [hep-th]].
%%CITATION = doi:10.1103/PhysRevD.78.023506, 10.1103/PhysRevD.79.129902;%%
%80 citations counted in INSPIRE as of 07 Jan 2020

	%\cite{Lehners:2010fy}
	\bibitem{Lehners:2010fy}
	J.~L.~Lehners,
	%``Ekpyrotic Non-Gaussianity: A Review,''
	Adv.\ Astron.\  {\bf 2010} (2010) 903907
	doi:10.1155/2010/903907
	[arXiv:1001.3125 [hep-th]].
	%%CITATION = doi:10.1155/2010/903907;%%
	%53 citations counted in INSPIRE as of 19 Jun 2019
	




%---------Ekp non-gau------- %


%\cite{Fertig:2013kwa}
\bibitem{Fertig:2013kwa}
A.~Fertig, J.~L.~Lehners and E.~Mallwitz,
%``Ekpyrotic Perturbations With Small Non-Gaussian Corrections,''
Phys. Rev. D \textbf{89} (2014) no.10, 103537
doi:10.1103/PhysRevD.89.103537
[arXiv:1310.8133 [hep-th]].
%40 citations counted in INSPIRE as of 06 May 2020




%\cite{Ijjas:2014fja}
\bibitem{Ijjas:2014fja}
A.~Ijjas, J.~L.~Lehners and P.~J.~Steinhardt,
%``General mechanism for producing scale-invariant perturbations and small non-Gaussianity in ekpyrotic models,''
Phys.\ Rev.\ D {\bf 89} (2014) no.12,  123520
doi:10.1103/PhysRevD.89.123520
[arXiv:1404.1265 [astro-ph.CO]].
%%CITATION = doi:10.1103/PhysRevD.89.123520;%%
%41 citations counted in INSPIRE as of 20 Dec 2019

%\cite{Boyle:2003km}
\bibitem{Boyle:2003km}
L.~A.~Boyle, P.~J.~Steinhardt and N.~Turok,
%``The Cosmic gravitational wave background in a cyclic universe,''
Phys. Rev. D \textbf{69} (2004), 127302
doi:10.1103/PhysRevD.69.127302
[arXiv:hep-th/0307170 [hep-th]].
%107 citations counted in INSPIRE as of 07 May 2020


%\cite{Hazumi:2019lys}
\bibitem{Hazumi:2019lys}
M.~Hazumi et al.,
%``LiteBIRD: A Satellite for the Studies of B-Mode Polarization and Inflation from Cosmic Background Radiation Detection,''
J. Low. Temp. Phys. \textbf{194} (2019) no.5-6, 443-452
doi:10.1007/s10909-019-02150-5
%45 citations counted in INSPIRE as of 07 May 2020

%---------Sphalerons------- %

%\cite{Kuzmin:1985mm}
\bibitem{Kuzmin:1985mm}
  V.~A.~Kuzmin, V.~A.~Rubakov and M.~E.~Shaposhnikov,
  %``On the Anomalous Electroweak Baryon Number Nonconservation in the Early Universe,''
  Phys.\ Lett.\  {\bf 155B} (1985) 36.
  doi:10.1016/0370-2693(85)91028-7
  %%CITATION = doi:10.1016/0370-2693(85)91028-7;%%
  %2712 citations counted in INSPIRE as of 11 Nov 2019


%---------Hypermagnetic fields------- %

%\cite{Turner:1987bw}
\bibitem{Turner:1987bw}
  M.~S.~Turner and L.~M.~Widrow,
  %``Inflation Produced, Large Scale Magnetic Fields,''
  Phys.\ Rev.\ D {\bf 37} (1988) 2743.
  doi:10.1103/PhysRevD.37.2743
  %%CITATION = doi:10.1103/PhysRevD.37.2743;%%
  %727 citations counted in INSPIRE as of 11 Nov 2019

%\cite{Garretson:1992vt}
\bibitem{Garretson:1992vt}
  W.~D.~Garretson, G.~B.~Field and S.~M.~Carroll,
  %``Primordial magnetic fields from pseudoGoldstone bosons,''
  Phys.\ Rev.\ D {\bf 46} (1992) 5346
  doi:10.1103/PhysRevD.46.5346
  [hep-ph/9209238].
  %%CITATION = doi:10.1103/PhysRevD.46.5346;%%
  %204 citations counted in INSPIRE as of 11 Nov 2019

%\cite{Joyce:1997uy}
\bibitem{Joyce:1997uy}
  M.~Joyce and M.~E.~Shaposhnikov,
  %``Primordial magnetic fields, right-handed electrons, and the Abelian anomaly,''
  Phys.\ Rev.\ Lett.\  {\bf 79} (1997) 1193
  doi:10.1103/PhysRevLett.79.1193
  [astro-ph/9703005].
  %%CITATION = doi:10.1103/PhysRevLett.79.1193;%%
  %267 citations counted in INSPIRE as of 11 Nov 2019
  
  
%\cite{Giovannini:1997eg}
\bibitem{Giovannini:1997eg}
  M.~Giovannini and M.~E.~Shaposhnikov,
  %``Primordial hypermagnetic fields and triangle anomaly,''
  Phys.\ Rev.\ D {\bf 57} (1998) 2186
  doi:10.1103/PhysRevD.57.2186
  [hep-ph/9710234].
  %%CITATION = doi:10.1103/PhysRevD.57.2186;%%
  %277 citations counted in INSPIRE as of 11 Nov 2019

	\bibitem{Giovannini:1999by}
	M.~Giovannini,
	%``Hypermagnetic knots, Chern-Simons waves and the baryon asymmetry,''
	Phys.\ Rev.\ D {\bf 61} (2000) 063502
	doi:10.1103/PhysRevD.61.063502
	[hep-ph/9906241].
	%%CITATION = doi:10.1103/PhysRevD.61.063502;%%


%\cite{Anber:2006xt}
\bibitem{Anber:2006xt}
  M.~M.~Anber and L.~Sorbo,
  %``N-flationary magnetic fields,''
  JCAP {\bf 0610} (2006) 018
  doi:10.1088/1475-7516/2006/10/018
  [astro-ph/0606534].
  %%CITATION = doi:10.1088/1475-7516/2006/10/018;%%
  %131 citations counted in INSPIRE as of 11 Nov 2019
	
	%\cite{Bamba:2006km}
	\bibitem{Bamba:2006km}
	  K.~Bamba,
	  %``Baryon asymmetry from hypermagnetic helicity in dilaton hypercharge electromagnetism,''
	  Phys.\ Rev.\ D {\bf 74} (2006) 123504
	  doi:10.1103/PhysRevD.74.123504
	  [hep-ph/0611152].
	  %%CITATION = doi:10.1103/PhysRevD.74.123504;%%
	  %26 citations counted in INSPIRE as of 11 Nov 2019
	  
%\cite{Martin:2007ue}
\bibitem{Martin:2007ue}
  J.~Martin and J.~Yokoyama,
  %``Generation of Large-Scale Magnetic Fields in Single-Field Inflation,''
  JCAP {\bf 0801} (2008) 025
  doi:10.1088/1475-7516/2008/01/025
  [arXiv:0711.4307 [astro-ph]].
  %%CITATION = doi:10.1088/1475-7516/2008/01/025;%%
  %214 citations counted in INSPIRE as of 11 Nov 2019

	  
%\cite{Demozzi:2009fu}
\bibitem{Demozzi:2009fu}
  V.~Demozzi, V.~Mukhanov and H.~Rubinstein,
  %``Magnetic fields from inflation?,''
  JCAP {\bf 0908} (2009) 025
  doi:10.1088/1475-7516/2009/08/025
  [arXiv:0907.1030 [astro-ph.CO]].
  %%CITATION = doi:10.1088/1475-7516/2009/08/025;%%
  %210 citations counted in INSPIRE as of 11 Nov 2019
  

%\cite{Durrer:2010mq}
\bibitem{Durrer:2010mq}
  R.~Durrer, L.~Hollenstein and R.~K.~Jain,
  %``Can slow roll inflation induce relevant helical magnetic fields?,''
  JCAP {\bf 1103} (2011) 037
  doi:10.1088/1475-7516/2011/03/037
  [arXiv:1005.5322 [astro-ph.CO]].
  %%CITATION = doi:10.1088/1475-7516/2011/03/037;%%
  %95 citations counted in INSPIRE as of 11 Nov 2019

%\cite{Caprini:2014mja}
\bibitem{Caprini:2014mja}
  C.~Caprini and L.~Sorbo,
  %``Adding helicity to inflationary magnetogenesis,''
  JCAP {\bf 1410} (2014) 056
  doi:10.1088/1475-7516/2014/10/056
  [arXiv:1407.2809 [astro-ph.CO]].
  %%CITATION = doi:10.1088/1475-7516/2014/10/056;%%
  %87 citations counted in INSPIRE as of 11 Nov 2019

%\cite{Adshead:2016iae}
\bibitem{Adshead:2016iae}
  P.~Adshead, J.~T.~Giblin, T.~R.~Scully and E.~I.~Sfakianakis,
  %``Magnetogenesis from axion inflation,''
  JCAP {\bf 1610} (2016) 039
  doi:10.1088/1475-7516/2016/10/039
  [arXiv:1606.08474 [astro-ph.CO]].
  %%CITATION = doi:10.1088/1475-7516/2016/10/039;%%
  %58 citations counted in INSPIRE as of 11 Nov 2019
	  
	  
%\cite{Kamada:2016eeb}
\bibitem{Kamada:2016eeb}
  K.~Kamada and A.~J.~Long,
  %``Baryogenesis from decaying magnetic helicity,''
  Phys.\ Rev.\ D {\bf 94} (2016) no.6,  063501
  doi:10.1103/PhysRevD.94.063501
  [arXiv:1606.08891 [astro-ph.CO]].
  %%CITATION = doi:10.1103/PhysRevD.94.063501;%%
  %33 citations counted in INSPIRE as of 11 Nov 2019


  
  \bibitem{Kamada:2016cnb}
  K.~Kamada and A.~J.~Long,
  %``Evolution of the Baryon Asymmetry through the Electroweak Crossover in the Presence of a Helical Magnetic Field,''
  Phys.\ Rev.\ D {\bf 94} (2016) no.12,  123509
  doi:10.1103/PhysRevD.94.123509
  [arXiv:1610.03074 [hep-ph]].
  %%CITATION = doi:10.1103/PhysRevD.94.123509;%%
  %31 citations counted in INSPIRE as of 13 Jan 2020
  
%\cite{Caprini:2017vnn}
\bibitem{Caprini:2017vnn}
  C.~Caprini, M.~C.~Guzzetti and L.~Sorbo,
  %``Inflationary magnetogenesis with added helicity: constraints from non-gaussianities,''
  Class.\ Quant.\ Grav.\  {\bf 35} (2018) no.12,  124003
  doi:10.1088/1361-6382/aac143
  [arXiv:1707.09750 [astro-ph.CO]].
  %%CITATION = doi:10.1088/1361-6382/aac143;%%
  %17 citations counted in INSPIRE as of 11 Nov 2019




%---------Intergalactic Magnetic Helicity------- %

%\cite{Tashiro:2013ita}
\bibitem{Tashiro:2013ita}
  H.~Tashiro, W.~Chen, F.~Ferrer and T.~Vachaspati,
  %``Search for CP Violating Signature of Intergalactic Magnetic Helicity in the Gamma Ray Sky,''
  Mon.\ Not.\ Roy.\ Astron.\ Soc.\  {\bf 445} (2014) no.1,  L41
  doi:10.1093/mnrasl/slu134
  [arXiv:1310.4826 [astro-ph.CO]].
  %%CITATION = doi:10.1093/mnrasl/slu134;%%
  %57 citations counted in INSPIRE as of 26 Nov 2019
  
  
  
%------------------ %
	
%\cite{Cook:2011hg}
\bibitem{Cook:2011hg}
J.~L.~Cook and L.~Sorbo,
%``Particle production during inflation and gravitational waves detectable by ground-based interferometers,''
Phys. Rev. D \textbf{85} (2012), 023534
doi:10.1103/PhysRevD.85.023534
[arXiv:1109.0022 [astro-ph.CO]].


\end{thebibliography}
\end{document}